\documentclass[final,3p,times,twocolumn]{elsarticle}
\usepackage{amsthm}
\usepackage{amssymb}
\usepackage{color,soul}
\usepackage{graphicx} 
\usepackage{epsfig}
\usepackage{lipsum}
\usepackage{natbib}
\usepackage{soul}
\usepackage{color}
\journal{Chinese Physics C}
\begin{document}

\begin{frontmatter}

\title{ Energy estimation of cosmic rays by air shower radio signals}
\author[1]{Fatemeh Latifian}

\ead{f.latifian@semnan.ac.ir}
\author[1]{ Gohar Rastegarzadeh\corref{cor1}}
\cortext[cor1]{corresponding author}
\affiliation[1]{organization={Faculty of physics, Semnan university},
	city={Semnan},
	country={Iran.}}

\ead{grastegar@semnan.ac.ir}

\begin{abstract}

 We present a method for reconstructing the primary energy of cosmic ray air showers using radio emission. The approach is based on  CoREAS simulations performed for both the SURA experiment and  a dense reference array of antennas. By comparing the simulated electric field intensity from the dense array with that obtained for the SURA configuration, we derive a scale factor $(C_{ij})$. This scale factor exhibits a correlation with the primary energy, decreasing as energy increases. Unlike previous approaches employed  by various experiments, our technique shows limited sensitivity to the shower core location. Even for the farthest core positions allowed by the array, the scale factor remains effectively independent of the core location. We also examined the accuracy and reliability of the reconstruction method. Finally, we reconstructed the primary energy of cosmic rays using our method with a maximum error of about 11\%  in our simulation tests.

\end{abstract}

\begin{graphicalabstract}
	\includegraphics[width=\textwidth,height=13cm]{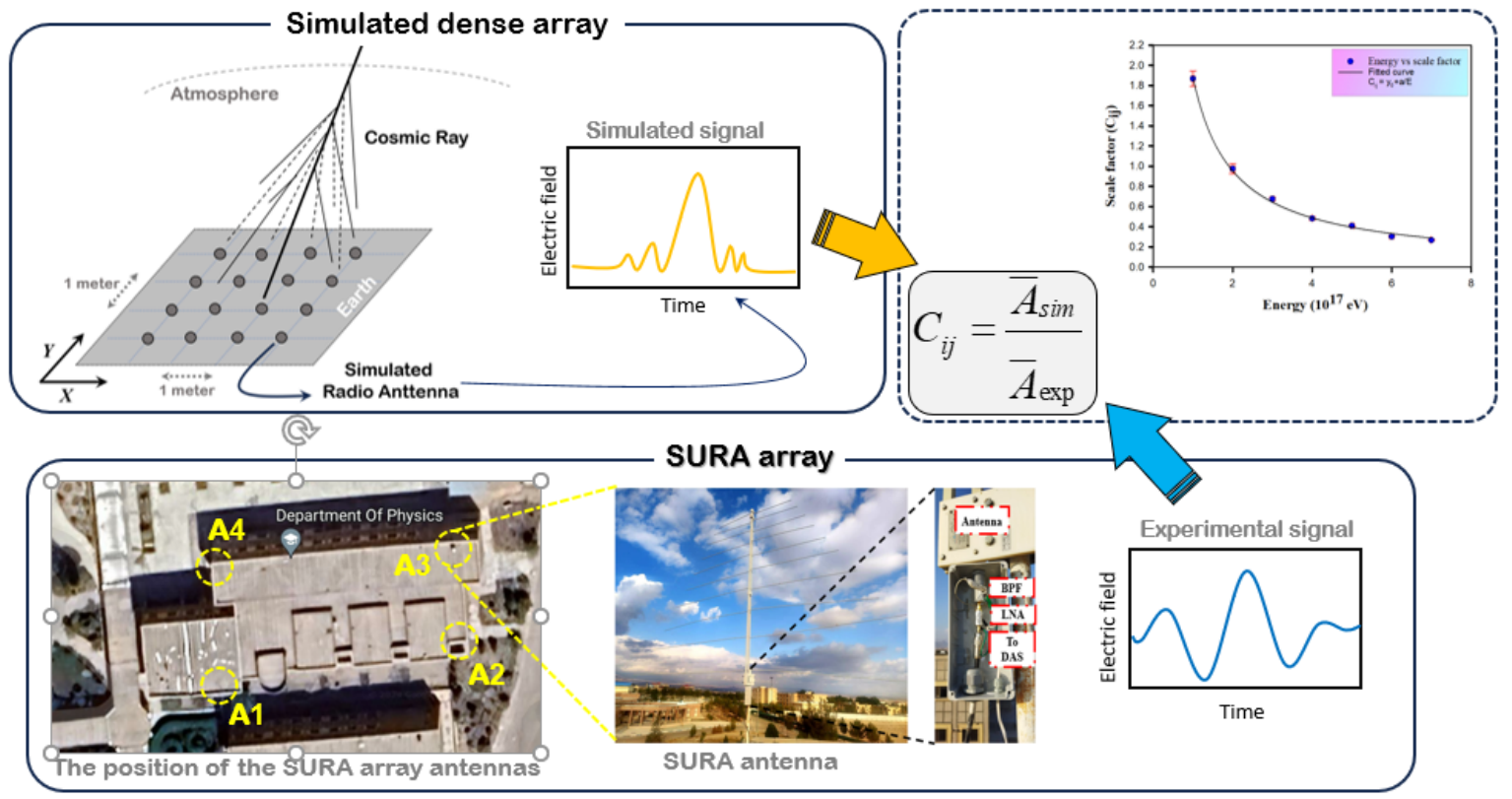}
\end{graphicalabstract}

\begin{highlights}
	\item The primary energy of cosmic rays can be reconstructed by comparing simulated and experimental radio signals
	
	\item The mean ratio between simulated and experimental radio signals (scale factor ($C_{ij}$)) is independent of the cosmic ray core locations

		\item There is a correlation between the scale factor and the primary energy of cosmic rays

\end{highlights}

\end{frontmatter}

\section{Introduction}	

\label{sec:intro}

 Cosmic rays are high-energy particles and photons originating from various sources outside the Earth's atmosphere, such as the sun, distant galaxies, supernovae, and other astrophysical phenomena. When these high-energy cosmic rays enter the Earth's atmosphere, they interact with nuclei of atoms they initiate cascades of secondary particles, resulting in what we call an Extensive Air Shower (EAS). As a result,  direct information about the primary particle's properties is largely lost. However, By detecting the secondary particles produced in the EAS and analyzing their Lateral Distribution Function (LDF), we can reconstruct some of the important characteristics of cosmic rays \cite{1}. For instance, it has been shown that the slope of LDF is proportional to the depth of shower maximum and therefore to the mass composition \cite{2}. In addition, the muon production depth distribution of secondary particles, has been widely employed to infer the mass composition of primary cosmic rays \cite{3}. In recent years, the detection of radio emissions from EAS has received significant attention due to its capability to determine essential parameters of cosmic rays. This approach offers several advantages, notably the ability to operate under various atmospheric conditions and its lower cost compared to other existing methods, such as Cherenkov and fluorescence light detection. These advantages have led multiple laboratories to adopt this approach alongside other techniques.
The SURA experiment is a radio array consisting of four log-periodic radio antennas located at Semnan University in Iran. By employing radio detection techniques, SURA has successfully reconstructed the arrival direction (zenith angle and azimuth angle) of cosmic rays, which is one of the key parameters of EAS using time information \cite{4}. Additionally, another method has been applied to determine the core location of cosmic ray air showers by comparing experimental and simulated signals \cite{5}. Another essential parameter of cosmic rays is the primary energy. The flux of cosmic ray energy decreases sharply as the primary energy increases, following a power-law distribution. Estimating the primary energy of cosmic rays is crucial for accurately determining the energy flux distribution. Reconstructing the primary energy of cosmic ray air showers is important as it contributes to our understanding of their origin, acceleration, and propagation mechanisms. AERA, Tunka-Rex, and LOPES experiments have utilized the LDF of cosmic ray air showers to determine the primary energy. These experiments have demonstrated that the intensity of radio signals at a specific distance from the core location of showers correlates with the primary energy, and each of these experiments have identified an optimum distance from the core position for their own radio station \cite{6,7,8}.  Moreover, the Antarctic Impulsive Transient Antenna (ANITA) experiment  was able to reconstruct the primary energy of cosmic ray air showers by using the off-axis angle of the Cherenkov ring  $\psi_{c}$ \cite{9}.  In  this paper, we propose a method to reconstruct the primary energy of cosmic rays at SURA, using a scale factor $(C_{ij})$. This scale factor $(C_{ij})$ compares the experimental radio signals with the simulated ones. This paper is structured into several sections: Section 2 provides an overview of the SURA experiment. Section 3 describes the calibration process, and Section 4 details the performed simulations. Moving on to section 5, we outline our method for determining the primary energy. In section 6, we discuss our results using our method, which is divided into three subsections, and in the final section we conclude our work.

\section{SURA: Semnan University Radio Array}

The Semnan University Radio Array, known as the SURA experiment, is composed of four  custom-built Log-Periodic Dipole Arrays (LPDA) oriented in the east-west direction and installed on the roof of the Physics Faculty at Semnan University, positioned at each corner. This configuration provides optimal coverage and maximizes the effective detection area, enhancing the array's ability to detect and analyze cosmic ray events. Each antenna consists of ten dipoles made from anodized aluminum alloy, chosen for their ultra-corrosion resistance. This material was specifically selected to withstand local weather conditions and wind speeds at the experiment's location. The dipoles are connected to a central waveguide positioned on top of each antenna. The LPDAs at SURA have linear polarization in the east-west direction, which is optimal for detecting cosmic ray signals, as the east-west component has the highest intensity compared to other components. At the SURA location, the average vertical atmospheric depth is 904 $gcm^{-2}$, with magnetic field components of $B_{x}$ = 39.53 $\mu$ T and $B_{z}$ = 28.1 $\mu$ T, which influence the propagation of cosmic rays and radio signals. To improve the quality of the detected signals, each antenna is equipped with a Low-Noise Amplifier (LNA) connected to the  antenna terminal, ensuring that even weak signals are captured effectively. A Band Pass Filter (BPF) is also used to remove undesirable radio emissions. The radio spectrum at the SURA array spans from 0 to 80 MHz, but due to the presence of substantial noise in the lower and upper parts of this range, particularly below 40 MHz and above 80 MHz, only the clear middle band is utilized. Consequently, the SURA array operates within the 40 to 80 MHz frequency range, where the signal-to-noise ratio is most favorable for detecting cosmic ray events. The raw frequency spectrum of the full 0 to 80 MHz range is shown in figure 1.
	
 	 	 \begin{figure}[!ht]
 		\centering
 		\label{fig:1}
 		\includegraphics[width=1\linewidth]{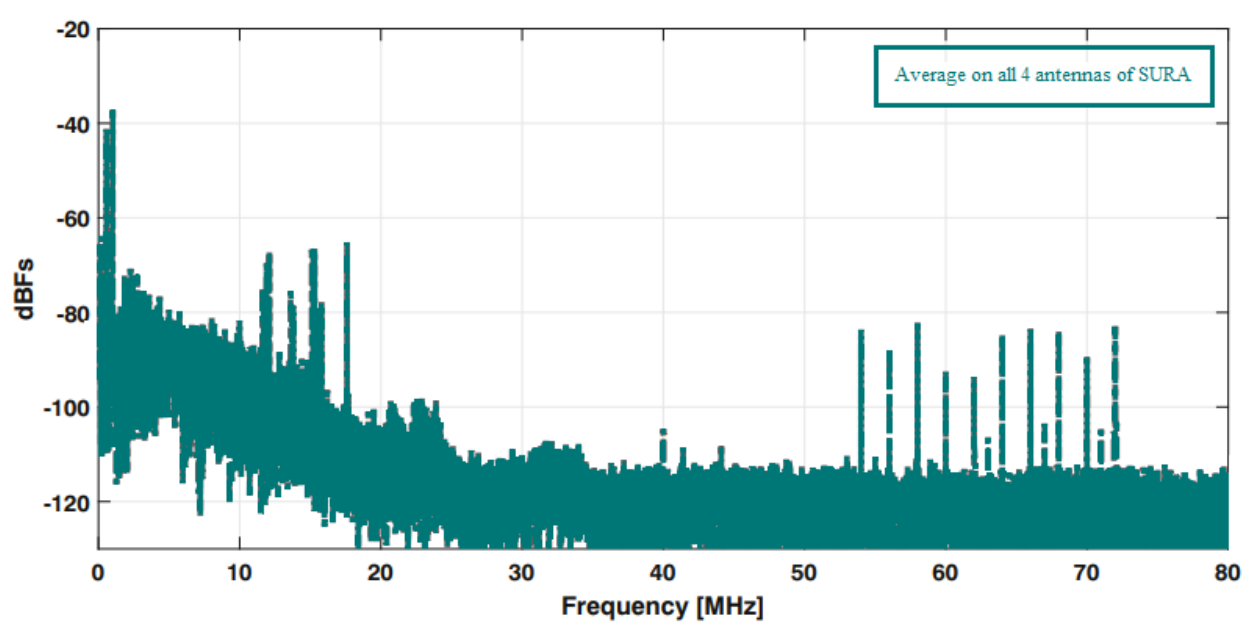}
 		\caption{ Frequency domain of the full spectrum, ranging from 0 to 80 MHz at the
 			SURA experiment.}
 	
 	\end{figure} 
 
 Based on our computer simulations and data analysis, along with an evaluation of background noise, we have determined that the SURA experiment is capable of detecting cosmic rays with energies above $10^{17}$ eV. The simulations confirm that the system's sensitivity is well-suited for capturing high-energy cosmic rays, making it a powerful tool for investigating these phenomena. The detected signals are then transferred to a  Data Acquisition System (DAS) using a high-quality RG213 cable  for further analysis and calibration \cite{10}. \ The DAS consists of a high-resolution Analog-to-Digital Converter (ADC) with a 160 MSps sampling rate and 14-bit resolution. Additionally the DAS includes an Intel File-Programmable Gate Array (FPGA) for online processing of the radio emissions. We investigated the frequency response of each device within the electronic chain of the SURA system. The overall system frequency response  was determined by examining the behavior of each component individually. In this process, we employed a signal generator to create radio signals, which were subsequently passed through each component of the SURA system, including the LNA, BPF, DAS, Cables, and connectors. These signals were then captured by an oscilloscope for analysis. By comparing the original signal adjusted for amplitude and frequency with the output signal passed through each device, we were able to evaluate the signal attenuation. This procedure allowed us to ascertain the frequency response and attenuation characteristics of each individual component within the system. We also utilized a computer simulation based on the ALTAIR FEKO package to analyze the behavior of the antennas. One of the SURA antennas is shown in Figure 2.

\begin{figure}[!ht]
	\centering
	\includegraphics[width=1\linewidth, height=0.8\columnwidth]{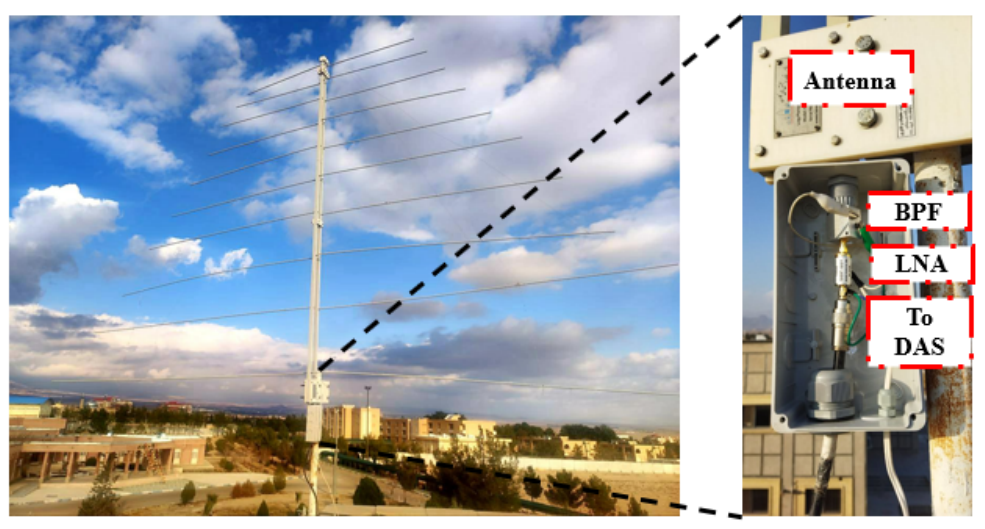}
	\caption{\label{fig:2} One of the SURA antennas.}

\end{figure}

\section{SURA data calibration}
 Data calibration in the SURA experiment is conducted in two primary stages to ensure the accuracy and reliability of the data. The first stage, real-time calibration, is managed by a Field Programmable Gate Array (FPGA) and carried out immediately as data is recorded. At this stage, only events that meet special quality criteria or exhibit desirable characteristics will be recorded and saved on a PC for further analysis and calibration. The second stage, post-processing calibration, involves additional refinements applied after data collection. These post-processing adjustments are implemented using a computer code developed specifically for this purpose.

\subsection{Real-time calibration via FPGA}
Real-time calibration is crucial for maintaining the integrity of the data during acquisition. Using the capability of the FPGA, we developed a computer program to process signals in real-time before saving them on a computer as a cosmic ray candidates. This program consists of several key components:

\begin{itemize}
	\item  \textit{Digital high-pass and band-pass filtering:} The program employs a digital high-pass and band-pass filter to reject unwanted radio emissions outside of the SURA frequency range, i.e.,  40 to 80 MHz.
	
	\item \textit{Notch filtering:}
As illustrated in Figure 1, there are a few mono-frequency noises within the operational frequency range of the SURA experiment, likely caused by TV or radio station activities. These noises are fixed in frequency and do not change over time. To eliminate these persistent mono-frequency noises, we employ several notch filters to effectively remove them. 
\item \textit{Threshold definition:}
	To ensure that only significant cosmic ray events are recorded, a threshold is also defined within the program. Given that the SURA antennas can  detect cosmic ray candidates with energies above $10^{17}$ eV, the code is configured to initiate data capture only for signals exceeding this energy threshold. This approach helps focus the data on events of interest, enhancing the relevance and quality of the recorded events.
	
	\item \textit{Coincident event investigation:} The program defines a specific time window during which any coincident events will be investigated.

	\subsection{post-processing calibration using a computer code}
After radio signals are recorded on a computer, we perform  post-processing calibration using a computer code to enhance the accuracy and reliability of the detected signals. The main steps in the post- processing calibration are:
	  
	\item \textit{Reconstruction of the arrival direction:}
	In the first step, the arrival direction of cosmic ray candidates is reconstructed with an angular accuracy of approximately $1^{\circ}$. This involves determining the zenith and azimuth angles by analyzing the timing of signals received by each antenna along and their spatial configuration.
	 
	\item \textit{ Applying the frequency response:}
	Based on the measurements detailed in \cite{10},  we determined the frequency responses of various components in the SURA  electronic chain, including the antennas, LNA, BPF, cables, and connectors. By applying the overall frequency response of the SURA array, we can accurately determine the intensity of the signals received by each antenna within the array.
	
	\item \textit{Frequency spectrum analysis:}
Despite the presence of band-pass filters (BPF) in the SURA electronic chain, we verify the frequency spectrum to ensure there are no unwanted signals.

   \item \textit{Filtering events based on azimuth angles:} In this step, we investigate the azimuth angle of detected events. Since many unwanted human-made signals are associated with large azimuth angles, we set a threshold to exclude events with azimuth angles beyond this limit, ensuring a cleaner dataset for further analysis.
  
	\item \textit{Wavelet noise reduction:}
	In this step, we apply the wavelet method for noise reduction of recorded events. By using the wavelet toolbox in MATLAB, we can decompose the signal into various frequency components and effectively isolate and reduce unwanted noises. This approach enhances the clarity of the data, allowing for more accurate analysis of radio signals.

	 \item \textit{Signal Up-Sampling:}
	In this step, we enhance the time resolution of the recorded signals by employing an up-sampling technique. The original signal resolution is determined by a time resolution of 6.25 ns. By implementing up-sampling, we improve the time resolution to 0.2 ns. 
	 
	 \item \textit{Signal Shape Filtering}
In the last step, we analyze the shape of the calibrated signals. By comparing the shape of calibrated radio signals to those expected from cosmic rays in our simulations, we filter out signals that do not match the characteristic cosmic ray profile, refining our dataset for further analysis.
\end{itemize}

\section{ CoREAS simulation setup} 
\label{sec:sim}

In this section, we simulate extensive air showers initiated by cosmic rays using the CoREAS extension of CORSIKA, which includes the full treatment of radio emission from electromagnetic cascades. The QGSJETII-04 model is employed for high-energy hadronic interactions, while  GHEISHA is used for low-energy  interactions, ensuring a consistent description of particle physics across the full energy rage of interest. We take into account all the essential features of the SURA experiment in the simulation, including the local magnetic field components and the height above sea level of 1130 meters, corresponding to the parameters of the SURA experiment, as both significantly influence the strength and polarization of the radio signals. The zenith angle at which most cosmic rays are received depends on three main factors: the solid angle element of the sky, the geometric efficiency of the detector, and the atmospheric depth at the observation level. Each radio experiment must carefully consider these factors to determine the optimal detection direction, as different experiments may require distinct zenith angles based on these factors. For the SURA experiment, analysis of the recorded cosmic ray candidates indicates that the orientation of the local magnetic field vector  ($\theta_{b}=36^{\circ}$ , $\phi_{b}=176^{\circ}$) together with the observation altitude leads to a maximum detection rate from showers arriving at a zenith angle of $35^{\circ}$ degrees.
As illustrated in figure 3, the distribution of detected events peaks at a zenith angle of $35^{\circ}$ which motivates our selection of this zenith angle for the simulations. For this zenith angle, the azimuthal distribution reaches at a maximum near $40^{\circ}$, which is also adopted as the selected azimuth angle.  

\begin{figure}[!ht]
	\centering
	\label{fig:3}
	\includegraphics[width=1\linewidth, height=0.7\columnwidth]{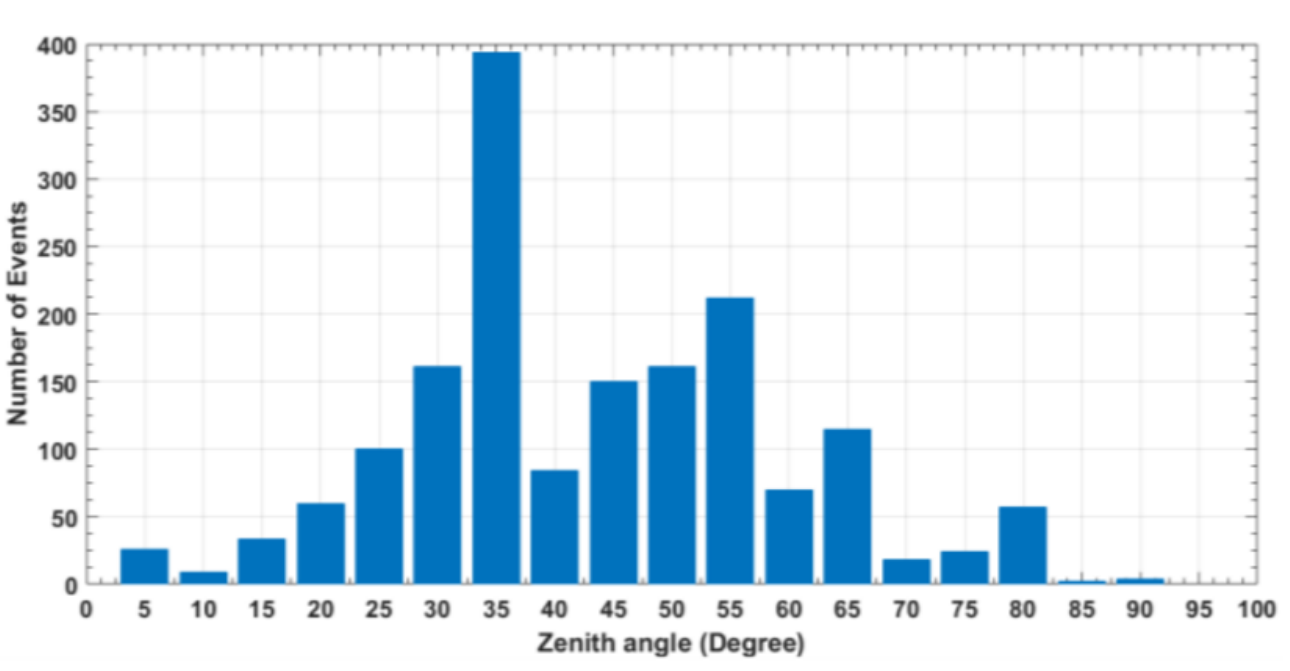}
	\caption{Distribution of zenith angle at SURA experiment.}
	
\end{figure} 

 Therefore, one of the arrival directions used in our study is the $35^{\circ}$ zenith angle and $40^{\circ}$ azimuth angle. However, to study the effect of arrival direction on our method, additional directions were also analyzed. We simulate seven sets of proton-induced showers with an arrival direction characterized by a zenith angle of 35$^{\circ}$  and an azimuth angle of 40$^{\circ}$, and with primary energy ranging from $10^{17}$eV to $7\times10^{17}$eV. Each set consists of 169 showers with different core locations along the SURA array. Totally, 1014 showers are simulated with this direction. Additionally, a dense array of 12342 radio antennas with one meter spacing is simulated to examine the signal simulation pattern. The schematic representation of a small part of the simulated dense array is shown in figure \ref{fig:2}, which illustrates the antenna layout and helps in understanding the distribution patterns within the simulation.  The dense array is simulated for a primary particle of a proton with 35$^{\circ}$ zenith angle and 40$^{\circ}$ azimuth angle, and primary energy of $2\times10^{17}$ eV, at core location ($x_{c}$=0,$y_{c}$=0). Although the dense array is simulated for this specific core, by using the virtual antenna method described in \ref{A}, we can use this array for investigating the cosmic ray primary energies for showers with other core locations. The characteristics of the dense array used in this study are summarized in Table \ref{tab:1}.  To study the effect of  arrival direction on energy reconstruction, approximately 525 showers are simulated with different core locations across two different zenith angles,  $60^{\circ}$ and $65^{\circ}$, both with the same azimuth angle of $40^{\circ}$. For each angle, two dense arrays consisting of about 400 radio antennas are set up , covering the region -125 m $<$x$<$125 m and -125 m $<$y$<$125 m at an energy of $2\times10^{17}$eV. Further details for each simulation are provided in the relevant sections.

\begin{table*}[!ht]
	\centering
	\caption{\label{tab:1} The dense array characteristics.}
	\begin{tabular}{c c c c c}
		\hline
		primary particle & Primary energy ( eV)& $\theta$ & $\phi$ & Number of antennas \\
		\hline
		Proton & 2 $\times$ $10^{17}$   & 35$^{\circ}$  & 40$^{\circ}$ & 12342\\
		
		\hline
	\end{tabular} 
\end{table*}

\begin{figure}[tbp]
	\centering
	\includegraphics[width=1\linewidth, height=0.8\columnwidth]{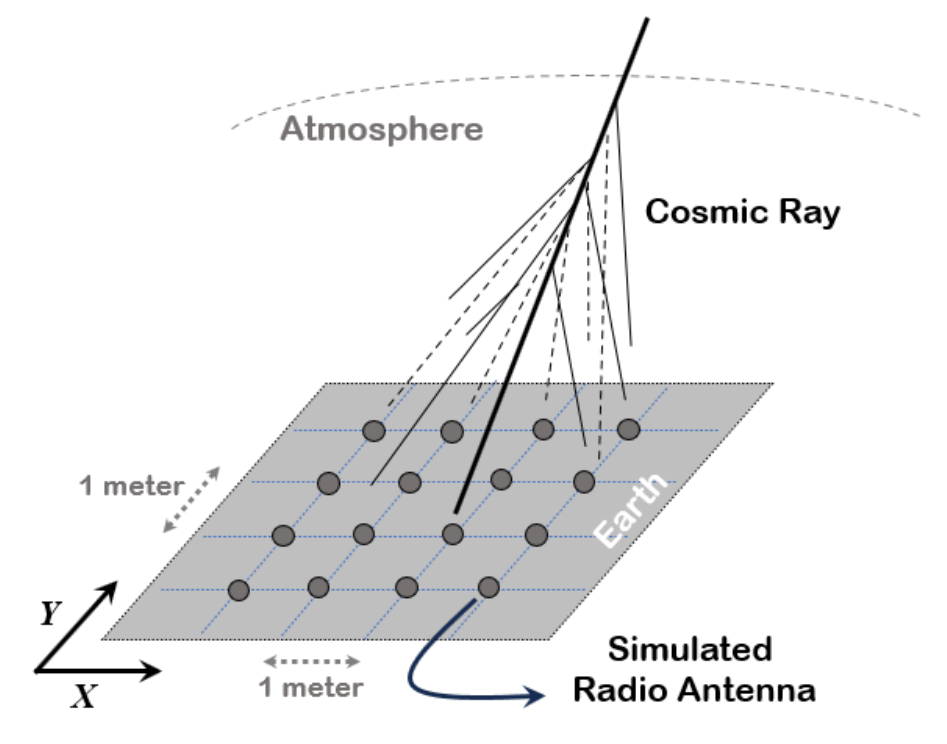}
	\caption{\label{fig:2} Schematic representation of a small part of the simulated dense array layout.}

\end{figure}

\section{Method of determining the primary energy}
\label{method}
In this section, we propose a method determine the primary energy of cosmic ray air showers by comparing simulated and experimental cosmic rays radio signals. We define a scale factor ($C_{ij}$) as the mean ratio between the peak intensity of experimental radio signals detected at the SURA experiment and the peak intensity of simulated radio signals for the dense array, which was described in section 3. This scale factor ($C_{ij}$) was previously introduced in the $\chi^{2}$ equation to estimate the core location of cosmic ray air showers \cite{11}.

\begin{equation}
	\label{eq:1}
	\chi_{ijk} ^2 = \chi^2 (x_{i},y_{j}) =\frac{1}{n} \sum_{k=1}^{n} \left[\frac{\frac{A^{sim}_{ijk}}{C_{ij}} - A^{exp}_{k} }{A^{exp}_{k}} \right]^2,
\end{equation}

\begin{equation}
	\label{eq:2}
	C_{ij}=\frac{\bar{A}^{sim}}{\bar{A}^{exp}}.
\end{equation}

In Equation \ref{eq:1}, $(x_{i}, y_{j})$ represents a core location being tested for the detected event, n is the number of antennas which, for instance, spans from 1 to 4 for SURA.  $A^{sim}$ represents the peak intensity of east-west components of simulated radio signals. While $A^{exp}$ denotes the peak intensity of east-west components of experimental radio signals. This By comparing the mean peak intensity of radio signals received by antennas at a radio array ($A^{exp}$) with their corresponding antennas at the dense array ($A^{sim}$), we can compute both the core location and the related scale factor for any recorded events. The tested guess $(x_{i}, y_{j}) $ for the core location that gives the minimum value to $\chi^{2}$ will be the acceptable core location. 
In this study,  $\lq\lq $ \emph{experimental radio signals}" refers to the  $\lq\lq $ \emph{simulated radio signals generated for the SURA experiment}". In future swork, we plan to use this technique for real signals detected by SURA.

\section{Result and discussion}
\label{Res}
 In this section, we discuss the results of applying the energy reconstruction technique described in this work to simulated EAS for the SURA experiment. The results are organized into three subsections. The first subsection examines the relationship between the scale factor and the primary energy. The second subsection evaluates the energy reconstruction error and proposed strategies to minimize it. The final subsection investigates the effect of the arrival direction on energy reconstruction.
 
\subsection{Applying the proposed method on simulated showers}

 In this subsection, we apply our method to simulated showers for the SURA experiment to reconstruct their primary energies. First, we demonstrate that the proposed method is largely  independent of the shower core location of cosmic ray air showers. Figure 5 shows the dependence of the scale factor on the core of cosmic ray air showers. Each plot in this  figure, represents 169 simulated air showers with an arrival direction characterized by a $35^{\circ}$ zenith angle and  a $40^{\circ}$  azimuth angle at various fixed primary energies but with different core locations within the range of -30 m $<x_{i}<$ 30 m and -30 m $<y_{i}<$ 30 m. The scale factors shown in figure 5 are obtained from minimizing the $\chi^{2}$ in  Equation 1.

\begin{figure*}[!ht]
	\centering
	\includegraphics[width=.33\textwidth,height=4.5cm]{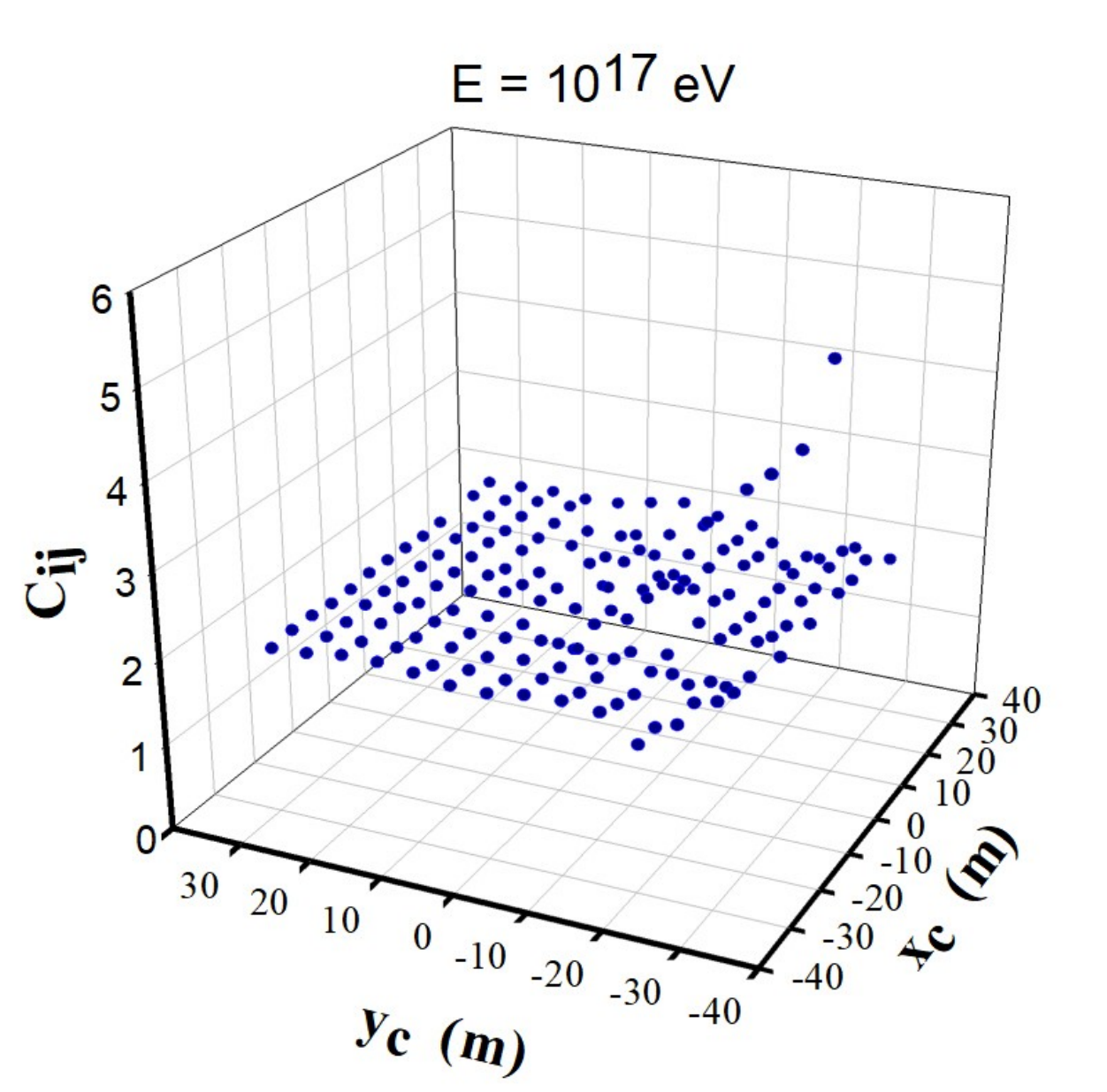}\hfill%
	\includegraphics[width=.33\textwidth,height=4.5cm]{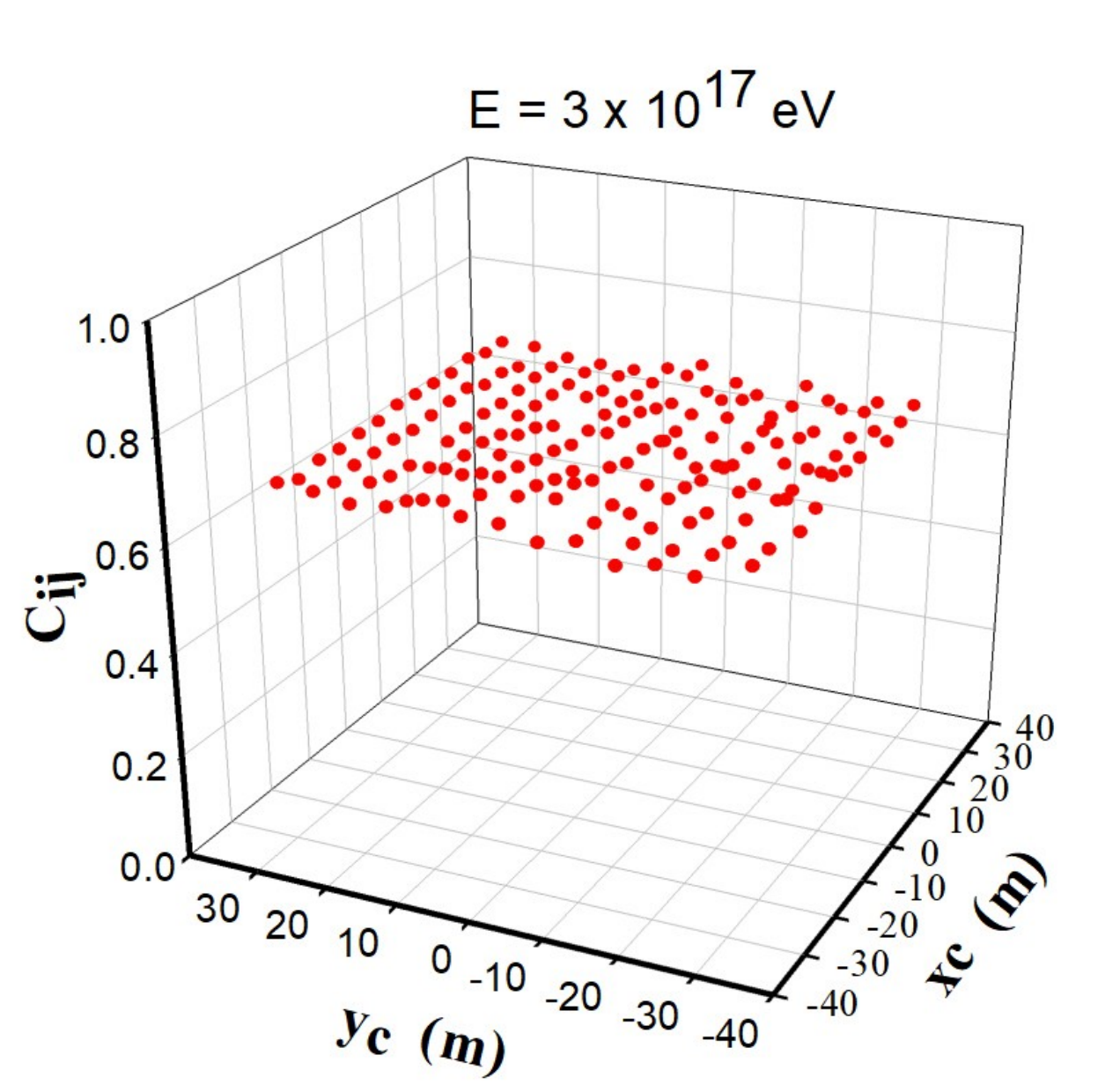}\hfill%
	\includegraphics[width=.33\textwidth,height=4.5cm]{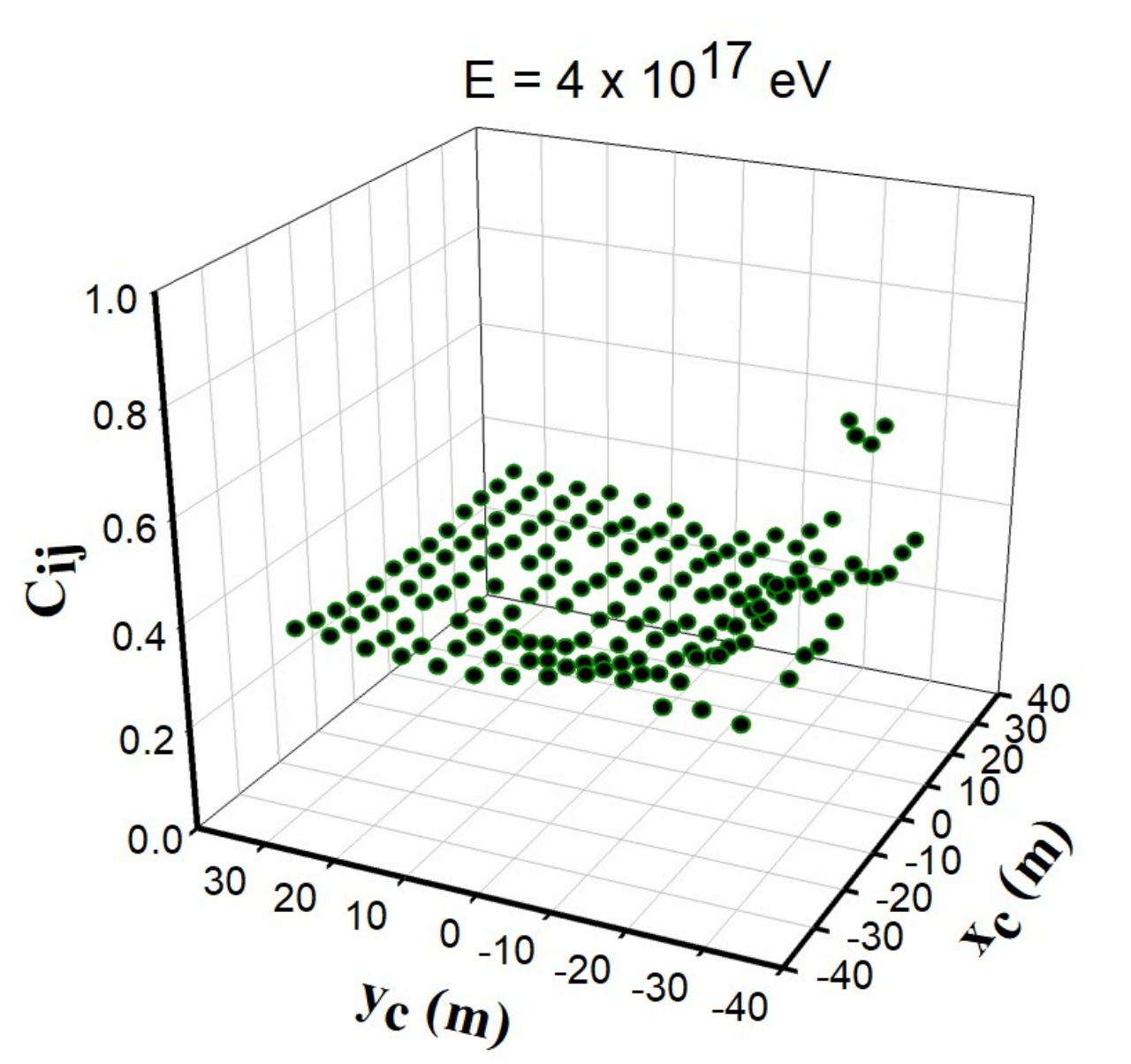}\\
	\includegraphics[width=.33\textwidth,height=4.5cm]{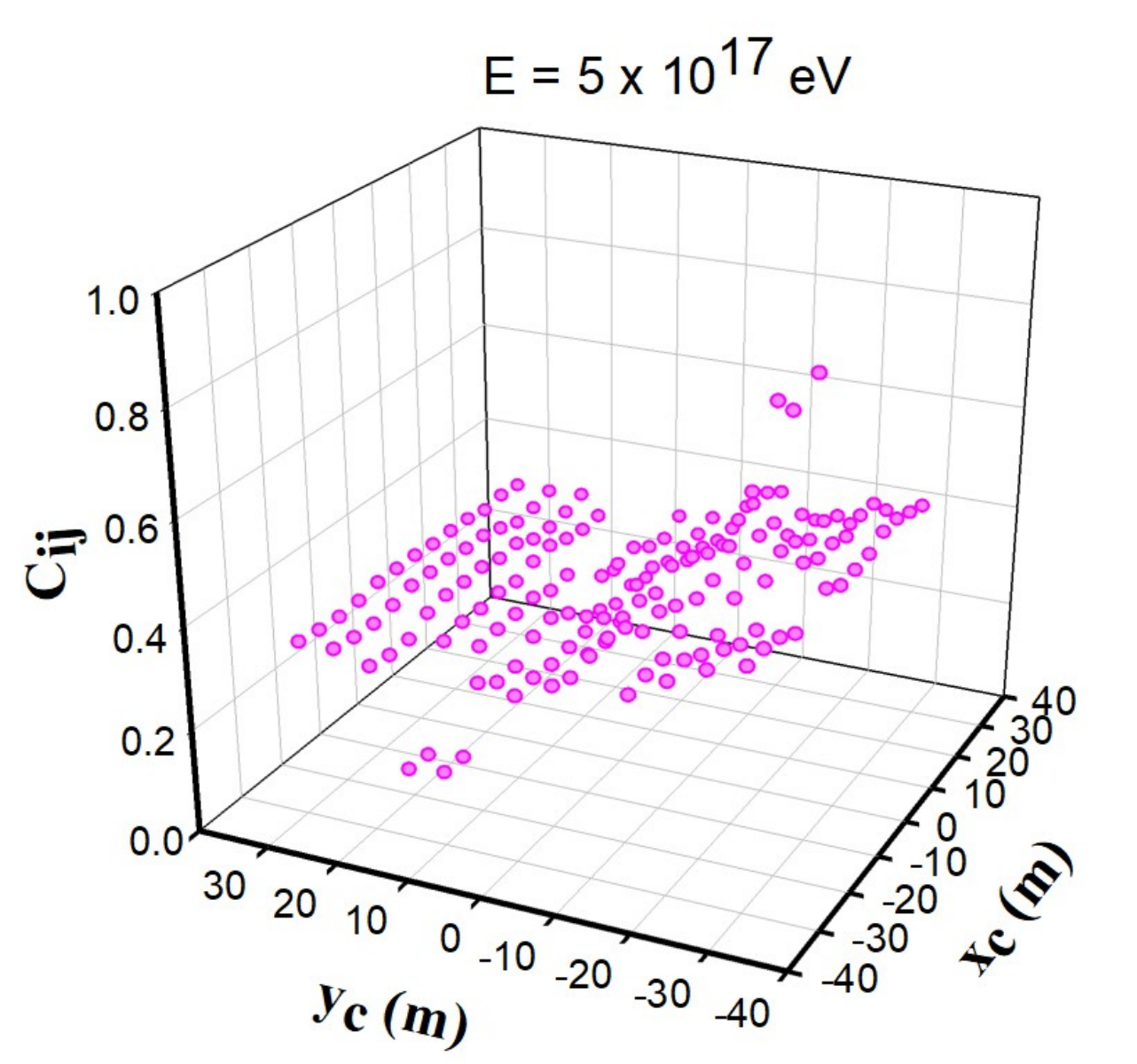}\hfill%
	\includegraphics[width=.33\textwidth,height=4.5cm]{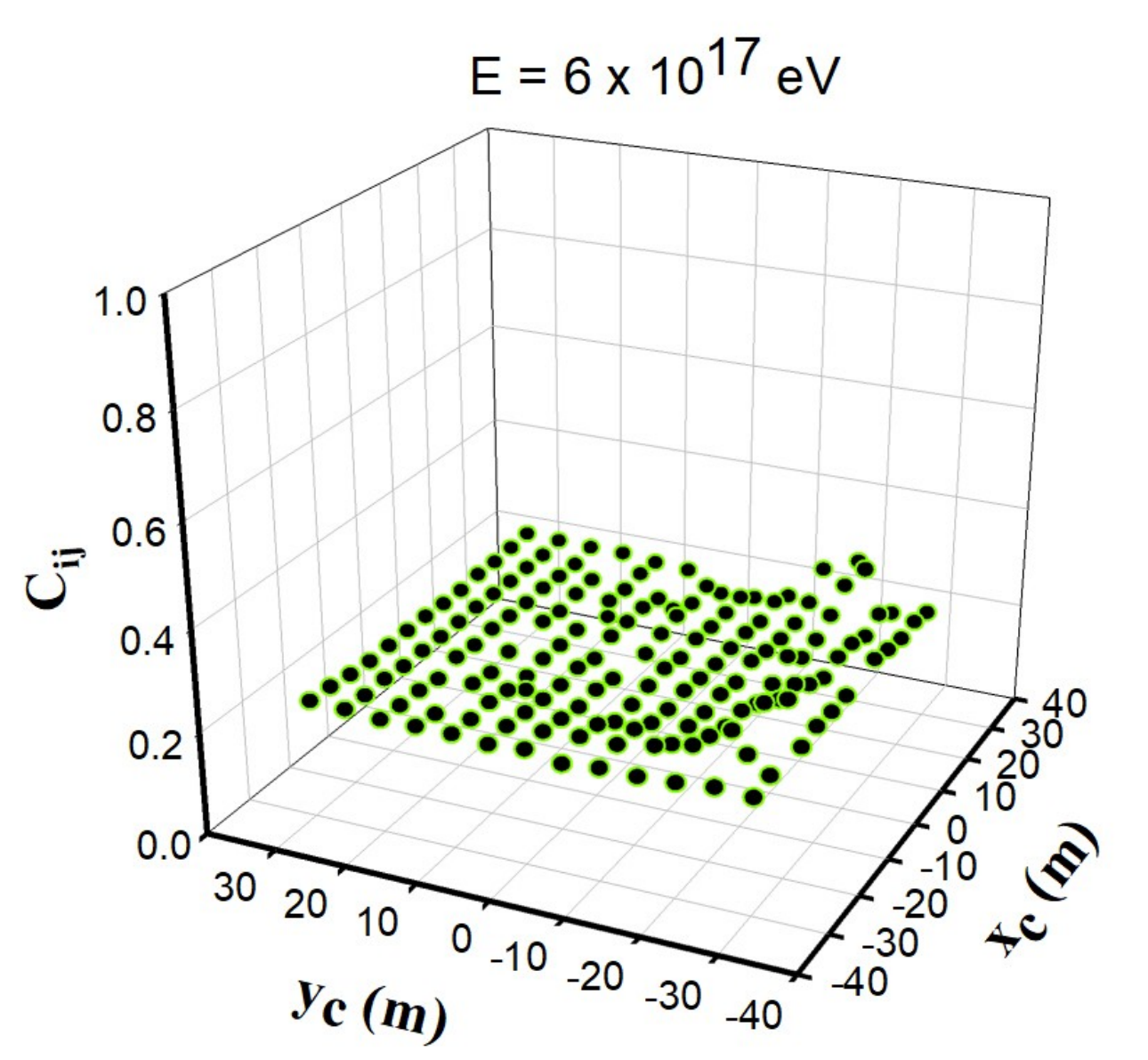}\hfill%
	\includegraphics[width=.33\textwidth,height=4.5cm]{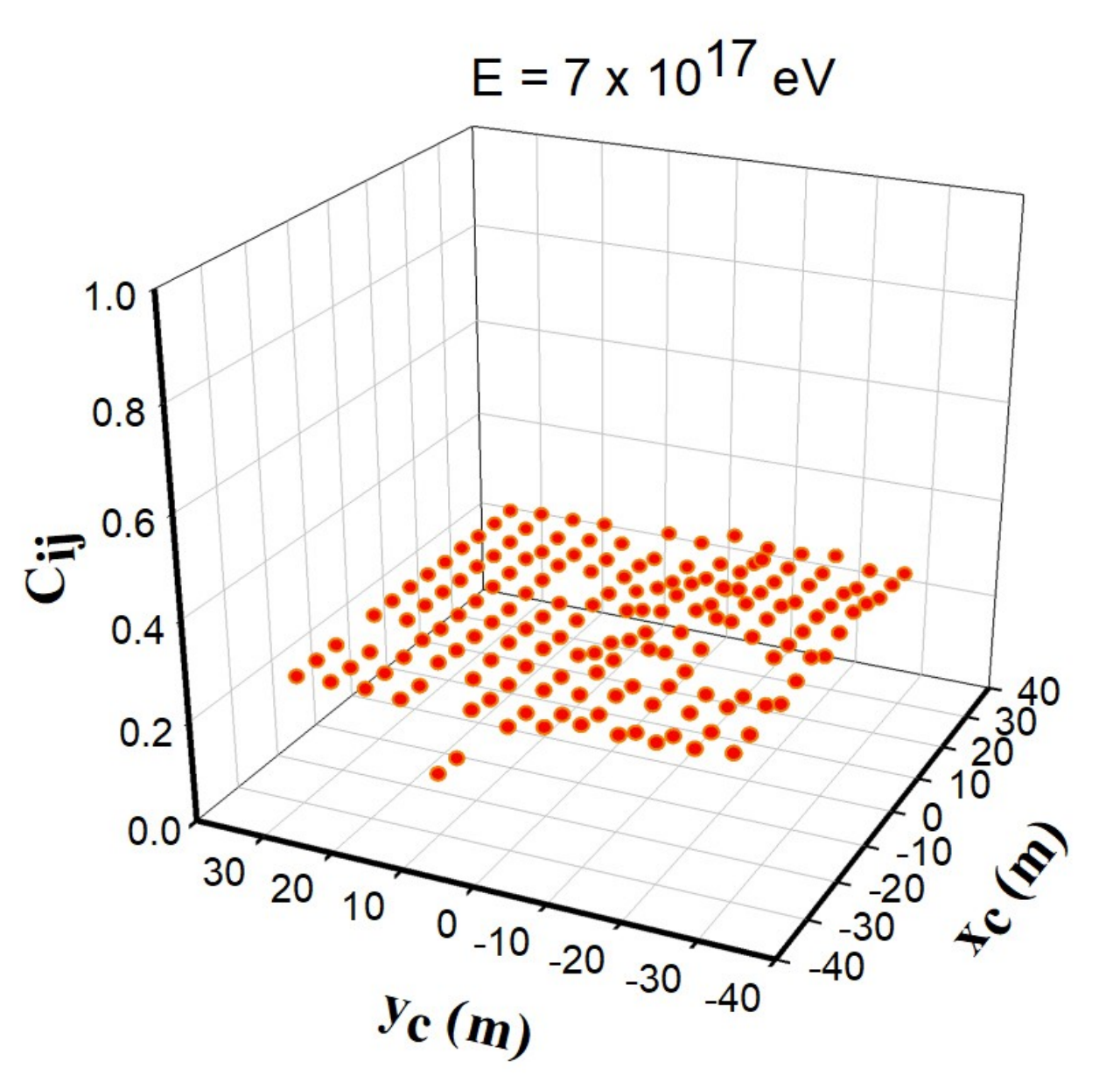}\\
	
	\caption{\label{fig:5}Scale factor $(C_{ij})$ versus Core location $(x_{c} ,y_{c})$ of 169 air showers.} 
	
\end{figure*}

As illustrated in figure~\ref{fig:5} it is evident that nearly all core locations lie on a plane at each fixed energy. In the next step, we will show that $(C_{ij})$  depends on the primary energy and can therefore be used to reconstruct the primary energy of EAS. To highlight the significance of the scale factor, we note that if the energy of an experimental event varies with the energy of the dense array, the intensity of the experimental signal will differ from the intensity of the signal received by the corresponding antenna in the dense array. To make these signals comparable, we use the scale factor. According to Equation \ref{eq:2}, if the primary energy of experimental events is smaller than that of simulated dense array shower, the scale factor will be greater than one. Otherwise, it will be smaller than one. When experimental and simulated showers have the same primary energy the scale factor will approach one. We calculate the mean scale factor of 169 simulated air showers at different energy levels, which is plotted in figure 6 (black dots). As shown in figure 6, as the primary energy of cosmic rays increases and deviates from the primary energy of simulated shower within the dense array ($2\times10^{17}$ eV), the intensity of the experimental signal ($A^{exp}$) exceeds the intensity of the simulated one ($A^{sim}$), resulting in a decrease in the scale factor. Each data point in this plot is accompanied by error bars indicating the uncertainty in both the energy and the scale factor. As we expected from the definition of the scale factor  ($C_{ij}=\frac{\bar{A}^{sim}}{\bar{A}^{exp}}$), the scale factor decreases with increasing primary energy and can be well described by an inverse first-order function of the form\newline

\begin{figure}[tbp]
	\centering
	\includegraphics[width=1\linewidth,height=0.8\columnwidth]{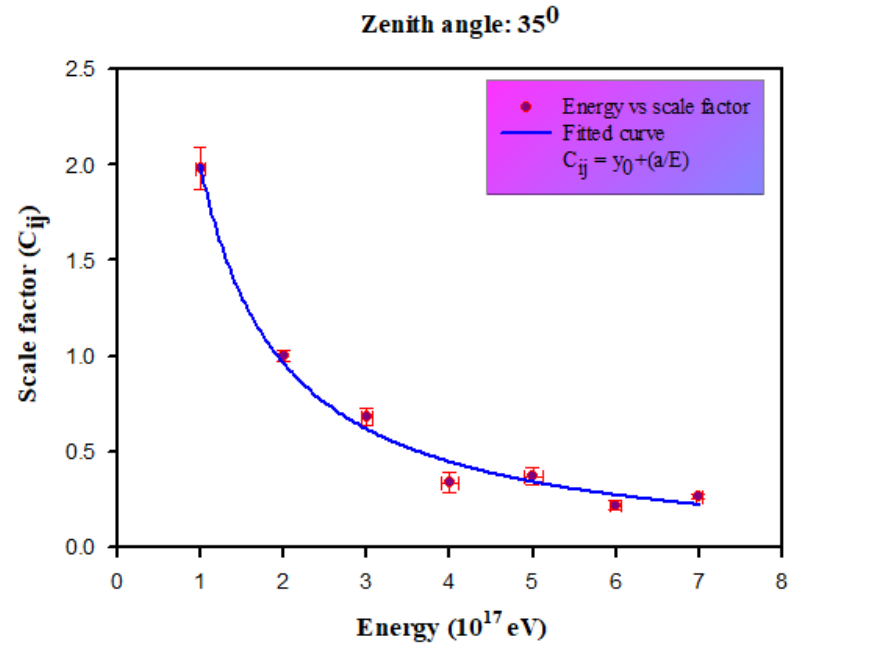}
	\caption{\label{fig:6} Correlation between the primary energy of cosmic ray air showers and the scale factor $(C_{ij})$ for showers with zenith angle of $35^{\circ}$.}
\end{figure}

\begin{equation}	
	C_{ij} = y_{0} +\frac{a}{E}  . 
\end{equation}

Where $y_{0}$ and a are the constant fitting parameters, with a = 2.070 and $y_{0}$ = -0.073 and E is the primary energy normalized by $10^{17}$ eV. That is, if the actual energy of a shower is 5$\times 10^{17}$ eV, the value to use in the equation is E = 5. This normalization allows convenient numerical representation of the fit without affecting the underlying physics. The inverse dependence of $(C_{ij})$  on the primary energy arises naturally from the well-established linear scaling of the radio electric field amplitude with primary energy as the primary energy increases the amplitude grows proportionally, while the scale factor decreases and asymptotically approaches a constant value. 
By determining the scale factor and 
 using Equation 3, the primary energy of cosmic rays can be estimated accurately,  the primary energy.

\subsection{Error analysis in Energy reconstruction }

 By differentiating Equation 3, the uncertainty in the reconstructed energy can be calculated as

\begin{equation}	
	\Delta E= - \frac{E^{2}  }{a} \Delta  C_{ij}.
\end{equation}

where $\Delta  C_{ij}$ is the standard deviation of the scale factor for a given set of showers. This uncertainty may be due to statistical fluctuations in the shower-to-shower variations of secondary particles and in the strength of their radio emissions.  From equation 4 we can also compute the relative error as
 
\begin{equation}
	\label{eq:5}	
	\frac{\Delta E}{E}= - \frac{E \times \Delta C_{ij} }{ a} \times 100 .
\end{equation}

For instance, the standard deviation of the scale factor ($\Delta c_{ij}$) for 169 simulated air showers at an energy level of  7 $\times$$ 10^{17}$ eV, turned out to be about 0.0136. By substituting this value along with constant parameter $\lq\lq $a"  into equation 5, we can compute the relative error of 4\% in determining the primary energy.
 Using this approach, we can calculate the relative error across various energy levels.  All reconstructed energies are shown in table \ref{tab:2}.
\begin{table*}
    \centering
	\caption{\label{tab:2} Comparison of   true and reconstructed primary energies with corresponding scale factors ($c_{ij}$) and error percentages at zenith angle of $35^{\circ}$.}	
	\begin{tabular}{c c c c}
		\hline
		True energy ($10^{17}$ eV)&reconstructed energy ($10^{17}$ eV)&$\Delta Cij$ & percentage of error  \\
		\hline
		1 & 1 $\pm$ 0.05 & 0.1078 & 5 \%\\
		2 &2 $\pm$ 0.05 &  0.0278 & 2  \%\\
		3 & 3 $\pm$ 0.18 & 0.0427& 6 \%\\
		4 & 4 $\pm$ 0.41& 0.0540& 10 \% \\
		5 & 5 $\pm$ 0.56 & 0.0467& 11 \% \\
		6 & 6 $\pm$ 0.41 & 0.0239& 6 \% \\
		7 & 7 $\pm$ 0.32 & 0.0136& 4 \%\\
		\hline
	\end{tabular} 
\end{table*}

\begin{figure}[tbp]
	\centering
	\includegraphics[width=1\linewidth,height=.81\columnwidth]{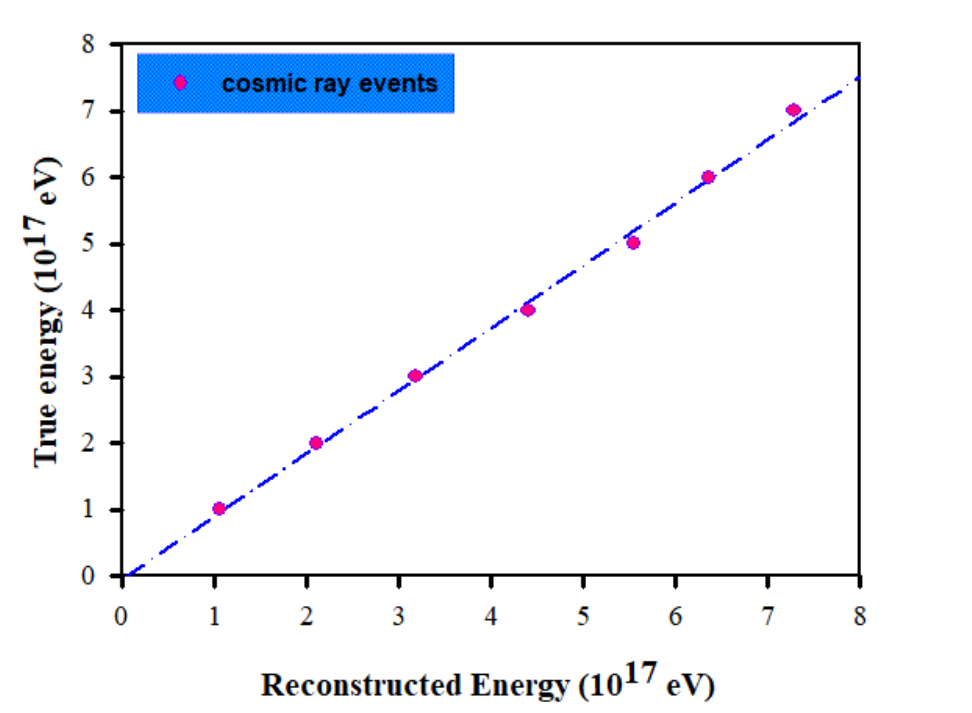}
	\caption{\label{fig:7} Comparison between reconstructed and true primary energies of cosmic ray air showers.}

\end{figure}

Figure \ref{fig:7} compares the true energy values from CoREAS simulation and reconstructed primary energy of cosmic rays using our method. It can be seen that the reconstructed  energies closely follow the true energies across the full energy range considered.

\subsection{The influence of arrival direction on energy reconstruction}
 Inclined cosmic ray showers have a larger lateral spread due to their extended paths through the atmosphere, leading to a significant footprint that affects the coherence of emitted radio signals. As the zenith angle increases, the footprint expands, while the intensity of the radio signals decrease, requiring simulations over a wider range of core locations to accurately assess energy reconstruction across different arrival directions. Due to the background noise at the SURA location, the array is capable of detecting cosmic rays with intensities greater than 200 $\mu V$. This threshold makes it challenging to detect inclined showers with low signal intensity. While this limits the detection of such showers at our current site, the approach could be adapted for quieter location or regions with lower background noise, where inclined showers would be more detectable. To investigate how different arrival direction impact the energy reconstruction of cosmic ray air showers, we simulated two additional zenith angles: $60^{\circ}$, $65^{\circ}$.  For each of these angles, we conducted simulations across seven energy levels, ranging from $10^{17}$ eV to 7 $\times$ $10^{17}$ eV. 
 At each energy level, 25 showers were simulated with different core locations  ranging from -125 m $<x_{i}<$ 125 m and -125 m $<y_{i}<$ 125 m, impacting the SURA experiment, with intervals of 50 meters, resulting in a total number of about 525 showers. We expanded the range of core locations beyond what was used in previous setups, allowing a more comprehensive exploration of how arrival direction effects the reconstruction accuracy. We also required two dense arrays at each arrival direction. Due to the expanded range of core locations compared to previous section, a larger dense array is required to cover the area. since simulating an extensive dense array is time-consuming, we optimized the setup by reducing the number of antennas within the dense array while expanding its overall coverage area. To address missing antennas, we applied 3D fitting to estimate signal intensities. Figure 8 illustrates the 3D fitting model used for the $60^{\circ}$ arrival direction to interpolate intensities across the array, highlighting the spatial variation of detected signal strength across different core locations.
 
 \begin{figure}[tbp]
 	\centering
 	\includegraphics[width=1\linewidth,height=0.95\columnwidth]{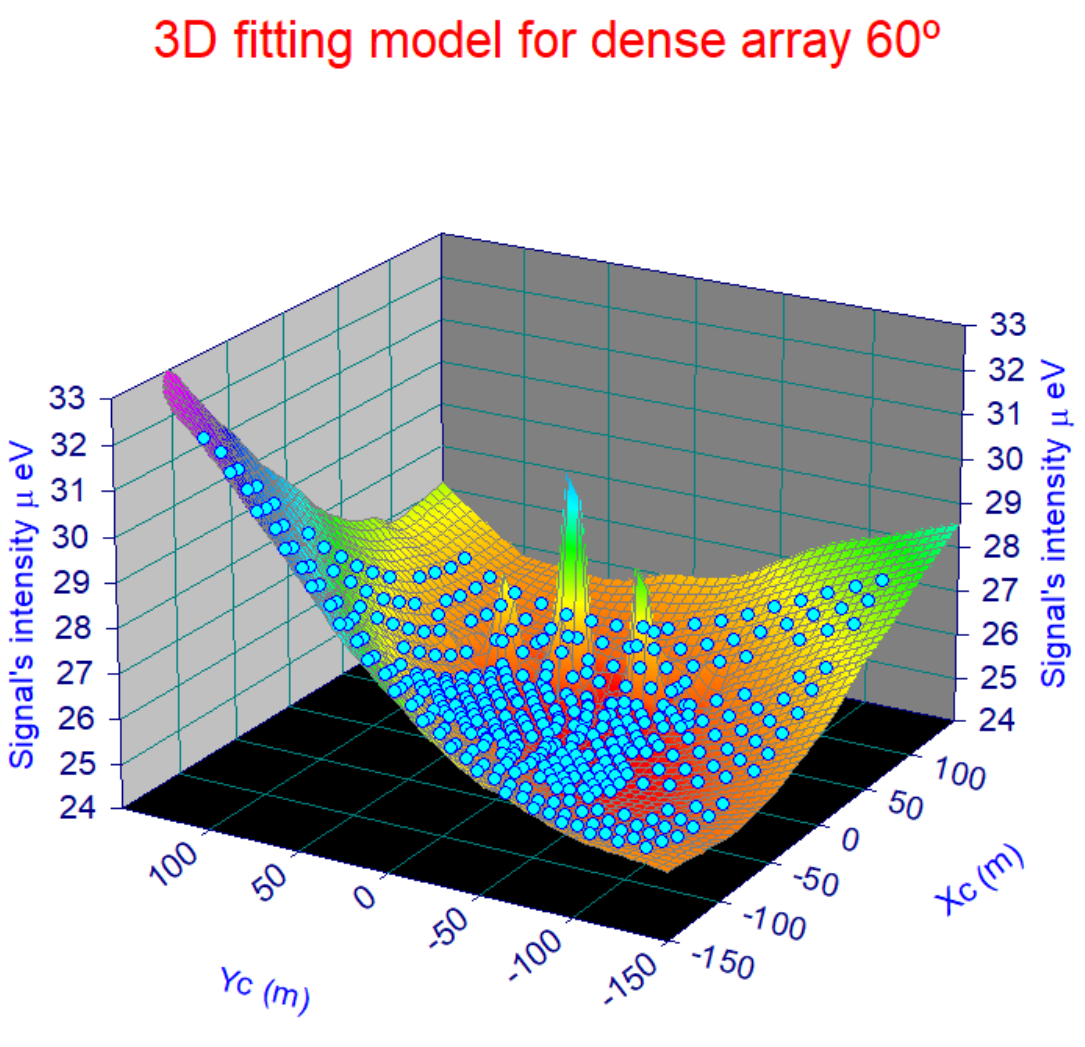}
 	\caption{\label{fig:6} 3D fitting model for dense array at $60^{\circ}$.}

 \end{figure}

  By comparing the peak intensity of radio signals simulated for SURA antennas,  $A^{exp}$, with those from the corresponding antennas in the dense array, $A^{sim}$, we calculate the scale factor for each shower.   For each arrival direction, we then determine the mean scale factor across showers simulated at various energies. Figure \ref{fig:6} presents the relationship between primary energy and the mean scale factor for three different arrival directions.

 \begin{figure}[!ht]
 	\centering
 	\includegraphics[width=.5\textwidth,height=7cm]{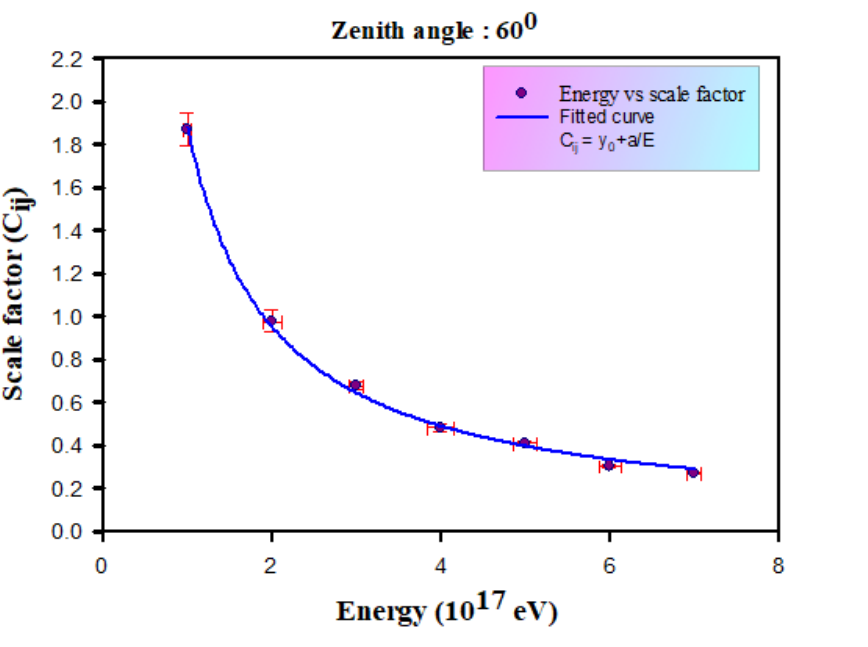}\hfill%
 	\includegraphics[width=.5\textwidth,height=7cm]{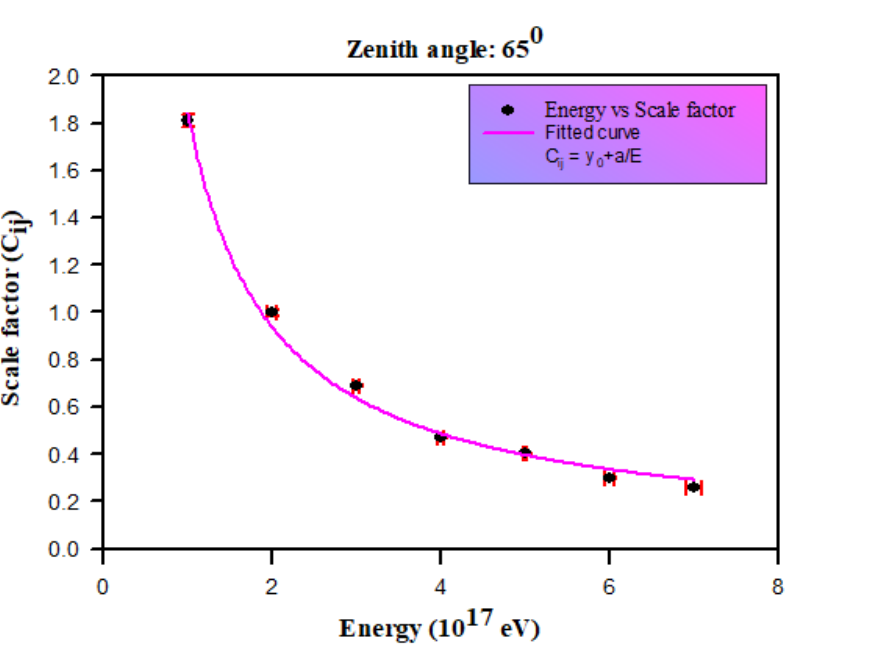}\\
 	\caption{\label{fig:6}The mean scale of showers at different energies and arrival directions.} 
 	
\end{figure}

 	 This figure allows for direct comparison of the scale factor trends across energies for each arrival direction, illustrating the influence of zenith angle on the energy reconstruction process. we observe  an inverse correlation between the scale factor and the primary energy of showers for each zenith angle, similar to the behavior observed in the $35 ^{\circ}$ case. The equation of the fitted curves are similar in all plots, differing only in constant parameters specific to each directions. The constant parameters for all equations are given in table 3. 
 	 
 	  \begin{table}[!ht]
 	  \label{tab:3}
 	  	\centering
 	  	\caption{ Constant fitting parameters for different directions.}
 	 \begin{tabular}{c c c c}
 	 	\hline
 	 	Zenith angle  & a & $y_{0}$  \\
 	 	\hline
 	 	$35^{\circ}$ & 2.07 & -0.073\\
 	 	$60^{\circ}$ & 1.86 &0.022\\
 	 	$65^{\circ}$ & 1.80& 0.033\\
 	 	\hline
	 \end{tabular} 
 	 \end{table}

 	 By using equation \ref{eq:5} and free parameters, we can calculate the error of energy reconstruction at all given directions using our method. To provide better insight and facilitate comparison between different zenith angles, we present the reconstruction errors in the table \ref{tab:4}. As can be seen, our proposed method is effective even for more inclined showers, confirming its validity across various zenith angles. we observed that the precision of our energy reconstruction is the lowest at the energy of 1 $\times$ $10^{17}$ eV and 5 $\times$ $10^{17}$ eV for the zenith angle of  $65^{\circ}$. Whereas, the maximum error is observed for the energy of 7 $\times$  $10^{17}$ for the zenith angle of $60^{\circ}$.

 	\begin{table*}[!ht]
 		\centering
 		\caption{\label{tab:4}  The error reconstruction of cosmic ray air showers at different energies and arrival directions.}
 		\label{tab:3}	
 		\begin{tabular}{c c c c c}
 			\hline
 			True energy ($10^{17}$ eV)& $35^{\circ}$ & $60^{\circ}$ & $65^{\circ}$ & \\
 			\hline
 			1 & 1 $\pm$ 0.05  &1 $\pm$ 0.04 & 1 $\pm$ 0.01&\\
 			2 &2 $\pm$ 0.05 & 2 $\pm$ 0.11 & 2 $\pm$ 0.06 & \\
 			3 & 3 $\pm$ 0.18 &  3 $\pm$ 0.08 & 3 $\pm$ 0.02 &\\
 			4 & 4 $\pm$ 0.41 & 4 $\pm$ 0.15 & 4 $\pm$ 0.03&\\
 			5 & 5 $\pm$ 0.56  &5 $\pm$ 0.13 &5 $\pm$ 0.01 &\\
 			6 & 6 $\pm$ 0.41&6 $\pm$ 0.12 & 6 $\pm$ 0.05 & \\
 			7 & 7 $\pm$ 0.32 & 7 $\pm$ 0.81 & 7 $\pm$ 0.09 &\\
 			\hline
 		\end{tabular} 
 	\end{table*}

\section{Conclusion}

We have presented  a  method to reconstruct the primary energy of cosmic rays using a scale factor ($C_{ij}$). The scale factor ($C_{ij}$) represents the mean ratio between the  peak intensity of simulated radio signals (signals from a simulated dense array) and the experimental data (here, simulated signals for SURA experiment). Initially, Based on CoREAS code we performed 1014 simulations of proton-induced showers across energy levels ranging from $10^{17}$ eV to $7\times 10^{17}$ eV, with an arrival direction characterized by zenith angle of $\theta$ = $35^{\circ}$  and azimuth angle of $\phi$ = $40^{\circ}$  for the SURA antennas. In addition, we simulated a dense array consisting of 12321 antennas, with the same arrival direction but with an energy level of  $2\times 10^{17}$ eV energy . Additionally, to investigate the influence of the arrival direction on determining the primary energy, we simulated more inclined showers and related dense arrays. Through a comparison between the signals simulated for the SURA at each antenna station and the signals received by their corresponding antennas in the dense array, we calculated the scale factor ($C_{ij}$) for each simulated shower. Afterwards, we demonstrated that there is an inverse correlation between the scale factor and the primary energy of cosmic rays. Finally, we could determine the primary energy of different cosmic ray events using this method with errors between 1\% to 11\%. We have demonstrated the effectiveness of our method even for inclined showers and extended core locations. In our future studies, once the SURA experimental data are completed, we aim to apply the method of reconstructing the primary energy to the experimental data detected by SURA as cosmic ray  candidates. Next, we aim to expand the size of array in simulations to simulate showers with more distant core locations and analyze more inclined showers. Finally, we intend to consider showers with other arrival directions to further test and validate the method.

\appendix
\section{The idea of virtual antenna}
\label{A}

\begin{figure}[!ht]
 	\centering
 	\label{3}
 	\includegraphics[width=1\linewidth,height=0.6\columnwidth]{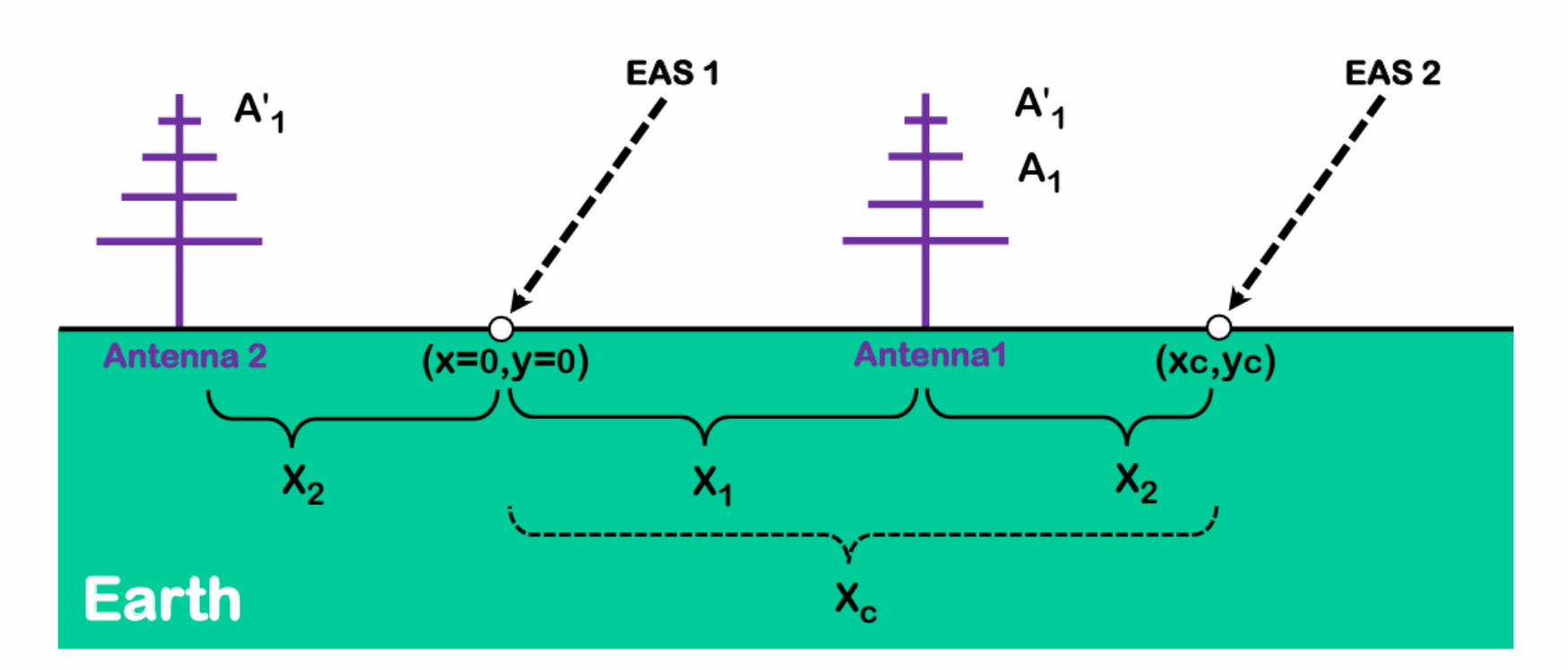}
 	\caption{\label{fig:1A} Virtual antenna.}
 \end{figure}

Radio antennas in a radio array detect different signal intensities depending on their distance from the shower core location. In figure \ref{fig:1A}, if the Extensive Air Shower number 1 (EAS 1) hits the center of the array, with core location of (x=0,y=0), with a specific  direction, antenna number 1 (A 1) will  receive the $E_{1}$ radio signal from this shower.
 On the other hand, if another air shower (EAS 2) strikes the array at the core location of $(x_{c},y_{c})$ with the same entry direction and the same primary mass and energy as the previous one, A 1 will receive a different signal than the previous one, which we call $E_{1}'$.
 
 Moreover, antenna number 2 (A 2), which is located at coordinates $x_{2}=x_{c}-x_{1}$ and $y_{2}=y_{c}-y_{1}$, receives the signal intensity $E_{1}'$ from EAS 1 with the core location at (x=0,y=0), which is equal to the signal intensity that A 1 receives from air shower with core location at ($x_{c},y_{c}$). Therefore, we call antenna number 2 the virtual antenna of antenna number 1. In other words, the signals received from air showers with core location of $(x_{c},y_{c})$ are identical to the signal received from air showers with core location of (x=0,y=0) by the virtual antenna corresponding to that antenna. Using this fact, we can utilize a dense array as a source of virtual antennas for investigating the cosmic ray primary energy which received at other core locations \cite{5}.

 \bibliographystyle{elsarticle-num-names} 
 \bibliography{bib}

\end{document}